\documentclass[prl,amsmath,twocolumn]{revtex4}  %% REVTeX 4.0
\usepackage{graphicx}% Include figure files

\begin{document}
\title{Oscillating tails of dispersion-managed soliton}

\author{Pavel M. Lushnikov$^{1,2,3}$
}

\affiliation{$^1$ Theoretical Division, Los Alamos National Laboratory,
  MS-B284, Los Alamos, New Mexico, 87545
  \\
  $^2$ Department of Mathematics, University of Arizona, PO Box 210089, Tucson,
  Arizona, 85721
  \\
  $^3$ Landau Institute for Theoretical Physics, Kosygin St. 2,
  Moscow, 119334, Russia
  }
%\email{lushnikov@cnls.lanl.gov}

%\date{Printed \today}

%\maketitle

\begin{abstract}
Oscillating tails of dispersion-managed optical fiber system are
studied for strong dispersion map in the framework of
path-averaged Gabitov-Turitsyn equation. The small parameter of
the analytical theory is the inverse time. An exponential decay in
time of soliton tails envelope is consistent with nonlocal
nonlinearity of Gabitov-Turitsyn equation, and the fast
oscillations are described by a quadratic law. The
pre-exponential modification factor is the linear function of time
for zero average dispersion and cubic function for nonzero
average dispersion.
\end{abstract}

% ocis{060.2330, 060.5530, 060.4370, 190.5530, 260.2030.}

\maketitle

%\begin{multicols}{2}

\section{Introduction}

A dispersion management is key current technology for development
of ultrafast high-bit-rate optical communication lines
\cite{kogelnik1,kurtzke1,chrapl1,smithknox1,gabtur1,gabtur2,kumar1,kaup1,mamyshev1,mamyshev2,turitsyn1,medvedev2002,gajadharsingh2004}.
A dispersion-managed (DM) optical fiber is designed to achieve
low (or even zero) path-averaged group velocity dispersion by
periodic alternation of the sign of the dispersion along an
optical line which dramatically reduces pulse broadening.

If nonlinearity is neglected it would be possible to achieve
transmission of optical pulses through DM system without
significant distortion. However, the linear transmission is
limited by the nonlinear distance $z<z_{nl}\equiv (\sigma
|u_0|^2)^{-1}$, which is determined by the Kerr nonlinearity
$\sigma$ and the characteristic pulse power $|u_{0}|^2$; $z$ is
the propagation distance along fiber. The characteristic power
can not be chosen too small in order to maintain appropriate
value of the signal-to-noise ratio.  As a result, Kerr
nonlinearity is essential for pulse propagation which results in
distortion of optical pulses and finite bit error rate (BER) for
information transmission. There are two alternative approaches to
deal with optical nonlinearity.

First approach is to reduce nonlinearity i.e. to make fiber
system as close as possible to linear system. E.g. one can use
low nonlinearity fiber~\cite{Hamaide95,Weins97} with increased
cross section. Another approach is to compensate nonlinearity
(nonlinearity management) using either semiconductor devices
~\cite{pare1996} or interferometer-based compensation of
nonlinear phase shift~\cite{GabitovLushnikov2002}. These systems
usually refer to as quasi-linear systems and they attracted very
much attention in recent years. Disadvantage of quasi-linear
systems is that nonlinearity is still essential at long enough
distance\cite{mamyshev1,mamyshev2} (e.g. transoceanic distance)
and it results in information losses. Also quasi-linear systems
often require complex and cumbersome dispersion
compensation\cite{mollenauer1tb2003}.

Second approach is not to reduce nonlinearity but rather to use
nonlineariy to achieve high bit rate transmission. This approach
is based on a concept of using of DM solitons  as the carrier of a
bit of information. DM soliton is a solitary pulse which
propagates in optical
fiber~\cite{smithknox1,gabtur1,gabtur2,kumar1,kaup1,mamyshev1,mamyshev2,turitsyn1}.
DM soliton experiences periodic oscillations, as a function of
propagation distance, with period of dispersion compensation.
Single DM soliton can propagate in optical fiber without
distortion (except periodic oscillations) for arbitrary long
distance even for nonzero path-averaged dispersion (dispersion,
averaged over period of dispersion variation). Propagation
without distortion results from balance between nonlinearity and
dispersion on the period of  dispersion compensation. In contrast,
quasi-linear pulse would increase its width with propagation
distance (in addition to periodic oscillations) due to nonzero
path-averaged dispersion.

DM soliton-based transmission has its own disadvantages caused by
interaction between different DM solitons (representing different
bits of information). There are two types of interactions. First
type is caused by the interaction between DM solitons with the
same carrier frequency (the same information channel). In
addition, modern optical lines use a wavelength-division
multiplexing (WDM) which allows the simultaneous transmission of
several information channels, modulated at different wavelengths,
through the same optical fiber. WDM means that there is also a
second type of interaction, which is the interaction between DM
solitons in different channels. Traditionally, the second type is
considered to be the most dangerous because DM solitons in
different channels move  with different group velocity and,
respectively, they pass through each other causing collisions
through nonlinear interaction. Collisions result in a jitter in
DM soliton arrival times and finite BER. However, recently, the
record-breaking, 1.09Tbits/s, transmission was
 achieved at 18,000 km optical line based on DM solitons~\cite{mollenauer1tb2003}
 which brought increasing interest to DM soliton-based transmission systems.
 The interchannel interactions in that transmission experiment were suppressed by a periodic-group-delay
 dispersion-compensating  module~\cite{mollenauerdelay2003}, and
 intrachannel interaction is most important. The main purpose of this Article
 is to determine the tails DM soliton which determines the interaction of DM solitons in the same channel.
 Soliton tails are responsible for interaction of pulses launched
into optical fiber which essentially limits bit-rate capacity of
optical line and makes finding DM soliton asymptotic behaviour an
important practical and fundamental problem.  It is found that
envelope of soliton decays exponentially in time (much slower
decay compare to Gaussian) so one can expect a strong interaction
of different DM solitons.

\section{Path-averaged equation}

 Neglecting polarization effects, higher order
dispersion, losses, stimulated Raman scattering and Brillouin
scattering, the propagation of optical pulse in a DM fiber is
described by a scalar nonlinear Schr\"odinger equation (NLS)
\begin{equation}
i A_Z -\frac{1}{2}\beta_2(Z)  A_{TT} +  \sigma(Z)|A|^{2} A =0,
\label{nls0}
\end{equation}
where $A$ is the envelope of optical pulse, $Z$ is the propagation
distance, $T$ is the  retarded time, $\beta_2(Z)$ is the
group-velocity dispersion which is the periodical function of
$Z$, $\sigma(Z)=(2\pi n_2)/\big (\lambda_0 A_{eff}(z)\big)$ is the
nonlinear coefficient, $n_2$ is the nonlinear refractive index,
$\lambda_0=1.55\, \mu m$ is the carrier wavelength, $A_{eff}$ is
the effective fiber area that in general case depends on $Z$.
Typical values of dispersion and nonlinear coefficient  are
$\beta_2=-20.0 \, \mbox{ps}^2/\mbox{km},$ $\sigma=0.0013 \,
(\mbox{km} \, \mbox{mW})^{-1}$ for standard monomode fiber, and
typical values for dispersion compensating fiber are
$\beta_2=103.9 \, \mbox{ps}^2/\mbox{km},$ $\sigma=0.00405\,
(\mbox{km} \, \mbox{mW})^{-1}$.

It is convenient to introduce the dimensionless variables:
$z=Z/l_0, \ t=T/t_0, \ u=A/\sqrt{P_0}$, where $l_0$ is the
typical period of dispersion variation, $t_0$ is the typical
pulse duration and $P_0$ is the typical pulse power. For typical
optical lines $l_0 \sim 50 \, \mbox{km}$, \ $t_0 \sim 10
\mbox{ps}, \ P_0\sim 1\, \mbox{mW}$. Eq. $(\ref{nls1})$ in
dimensionless variables takes the following form:
\begin{equation}
i u_z +   d(z) u_{tt} +  c(z)|u|^{2} u =0,  \label{nls1}
\end{equation}
where $c(z)=\sigma(z) P_0l_0, \ d(z)=-\beta_2l_0/(2t_0^2)$.
Consider a two-step periodic dispersion map (see Fig. 1):
$d(z)=\langle d\rangle+\tilde d(z)$, where $\tilde d(z)=d_1$ for
$0<z+n L<L_1$ and $\tilde d(z)=d_2$ for $L_1<z+n L<L$, $L\equiv
L_1+L_2$ is the dispersion map period, $\langle d \rangle$ is the
path-averaged dispersion, $d_1, d_2$ are the amplitudes of
dispersion variation subjected to condition $d_1L_1+d_2L_2\equiv
0$ and $n$ is an arbitrary integer number. Path averaged
dispersion is assumed to be small, $|\langle d \rangle|\ll |d_1|,
\, |d_2|.$

\begin{figure}%[htbp]
\begin{center}
\includegraphics[width = 3.5 in]{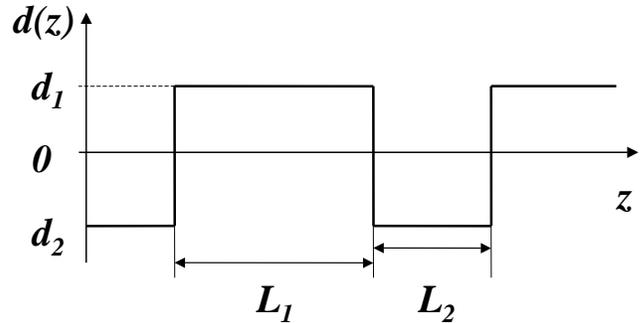}
%\vspace{0.5cm}
\caption{Dispersion $d(z)$ as function of propagation distance
$z$. } %\label{fig:fig1}
\end{center}
\end{figure}

Optical pulse experiences strong oscillation as a function of $z$
on the period of dispersion map $L$. These oscillations are caused
primarily by linear effects which are described by the linear part
of Eq. $(\ref{nls1})$.

A typical power of pulse is about a few $mW$ in most optical
fiber systems. As a result the nonlinearity in Eq. $(\ref{nls1})$
can be treated as a small perturbation on scales of typical
dispersion map period $L\sim 50km$ because the characteristic
nonlinear length $Z_{nl}$ of the pulse is large: $Z_{nl}\gg L,$
where $Z_{nl}=1/|p|^2$ and $p$ is the typical pulse amplitude. As
nonlinearity is small perturbation on distance $L$, it is
convenient to introduce the new variable$\psi:$ $\hat\psi \equiv
\hat u e^{i\omega^2\int^z_{L_1/2} \tilde d(z')dz'}$. Here
$\hat\psi(\omega)=\int^\infty_{-\infty}\psi(t)e^{\imath \omega
t}dt$ is  Fourier transform  of $\psi$ and similar to $\hat u$.
$\psi$ is the slow function of $z$ on the scale $L$ because all
fast linear dependence on $z$ is included into fast varying
exponent $e^{-i\omega^2\int^z_{L_1/2} \tilde d(z')dz'}$.

Fourier transform of Eq. $(\ref{nls1})$ gives:
\begin{eqnarray} \label{psiint0}
i \hat \psi_z(\omega) -\omega^2 \langle d \rangle \hat \psi
+\frac{c(z)}{(2\pi)^2}\int
\exp\Big [-i\triangle\int^z_{L_1/2} \tilde d(z')dz'\Big ]   \nonumber \\
\times\hat\psi(\omega_1)\hat\psi(\omega_2)\hat\psi^*(\omega_3)
\delta(\omega_1+\omega_2-\omega_3-\omega) d\omega_1 d\omega_2
d\omega_3,
\end{eqnarray}
where $\triangle\equiv\omega_1^2+\omega_2^2-\omega_3^2-\omega^2$.
Because $\hat \psi$ is the slow function one can approximate its
derivative as $\hat\psi(z)_z\simeq [\hat
\psi(z+L)-\hat\psi(z)]/L$ and integrate Eq. $(\ref{psiint0})$
over dispersion map period $L$ neglecting  the dependence of
$\psi$ on $z$ in the nonlinear term. This averaging procedure
results in path-averaged Gabitov-Turitsyn \cite{gabtur1,gabtur2}
model:
\begin{eqnarray} \label{psiint1}
i \hat \psi_z(\omega) -\omega^2 \langle d \rangle \hat \psi
+\frac{c_0}{(2\pi)^2}\int
\frac{\sin{\frac{s\triangle}{2}}}{\frac{s\triangle}{2} } \hat\psi(\omega_1)\hat\psi(\omega_2)  \nonumber \\
\times\hat\psi^*(\omega_3)
\delta(\omega_1+\omega_2-\omega_3-\omega) d\omega_1 d\omega_2
d\omega_3, \nonumber \\
c_0=(c_1L_1+c_2 L_2)/L,
\end{eqnarray}
where $s=d_1 L_1$ is the dispersion map strength, and $c_1,\ c_2$
are the values of $c(z)$ for $0<z+n L<L_1$ and  $L_1<z+n L<L$,
respectively. It is set below, without loss of generality, that
$c_0=1$ because one can always define typical power $P_0$ in such
a way to insure $c_0=1.$ Note that Gaitov-Turitsyn Eq. can be also
obtained by averaging of the Hamiltonian, $H=\int\big [
d(z)|\psi_t|^2-|\psi|^4/2\big ] dx$, of NLS $(\ref{nls1})$ over
dispersion map period \cite{gabtur2}. The numerical solutions of
the Gabitov-Turitsyn model were compared with simulations of full
NLS $(\ref{nls1})$ which showed good agreement
\cite{turitsyn1,ablowitz1}. Returning to time domain in Eq.
$(\ref{psiint1})$ one gets \cite{ablowitz1}:
\begin{eqnarray} \label{psiint1b}
i \psi_{ z} + \langle d \rangle \psi_{tt}- \frac{1}{2\pi s}\int
Ci(\frac{t_1 t_2}{s}) \psi(t_1+t)\psi(t_2+t)\times \nonumber
\\ \psi^*(t_1+t_2+t) dt_1 dt_2=0,
\end{eqnarray}
where $Ci(x)=\int^x_\infty\frac{\cos{x}}{x}dx$ (note difference in
definition of $Ci(x)$ in comparison with \cite{ablowitz1}).

Equation for DM soliton solution, $\psi= A(t)e^{i\lambda z} $ ($A$
is real), of the Gabitov-Turitsyn Eq. $(\ref{psiint1b})$  takes
the following form:
\begin{eqnarray} \label{psiint1c}
-\lambda A + \langle d \rangle A_{tt}= \frac{1}{2\pi s}\int
r(t_1,t_2,t) dt_1dt_2, \nonumber \\
 r(t_1,t_2,t)=Ci\big(\frac{t_1 t_2}{s}\big )
A(t_1+t)A(t_2+t)A(t_1+t_2+t),
\end{eqnarray}
where $Ci(x)=\int^x_\infty\cos{x}/xdx$.

Eq. $(\ref{psiint1c})$ is used throughout this Article Eq. to
study DM soliton.

\section{DM soliton tails}

The previous studies
\cite{kurtzke1,chrapl1,smithknox1,gabtur1,gabtur2,kumar1,kaup1,turitsyn1,pelin2000,zharnitsky1,zharnitsky2,Kunze2003,lush2000a,lush2000b,lush2001a}
have essentially focused on finding of DM soliton width,
amplitude and soliton shape near soliton center. It was found
that Gaussian ansatz,
\begin{equation} \label{psi0}
A_{Gauss}=p \exp{\big ( -\frac{\beta}{2}t^2\big )},
\end{equation}
where $p, \beta$ are real constants,
 is a rather good approximation for the DM
soliton solution near soliton center \cite{smithknox1}. Solution
of Eq. $(\ref{psiint1c})$ by iterations allows to find soliton
amplitude and width analytically with accuracy $\sim 1\%$ (see
Refs. \cite{lush2000a,lush2001a}). Note that parameters $s, \,
\langle d\rangle, \, \lambda$ uniquely determine the numerical
form of DM soliton \cite{lush2001a}. E.g. for $\langle
d\rangle=0$, $\beta s=2.393\ldots,\ \lambda=p^2 \times 0.482
\ldots$ \cite{lush2000a}. Thus dispersion map strangth, $\beta
s$, is fixed and not small for DM soliton. However an important
question about asymptotic behavior of soliton tail for large time
$t$ was only marginally addresses in the past
\cite{pelin2000,lush2001a}. Soliton tails are responsible for
interaction of pulses launched into optical fiber which
essentially limits bit-rate capacity of optical line and makes
finding DM soliton asymptotic behaviour an important practical
and fundamental problem. DM soliton tails behaviour is the main
subject of this Article.

Solid curves in Fig. 2a and Fig. 2b show
\begin{figure}%[htbp]
{\bf (a)}
\begin{center}
\includegraphics[width = 3.5 in]{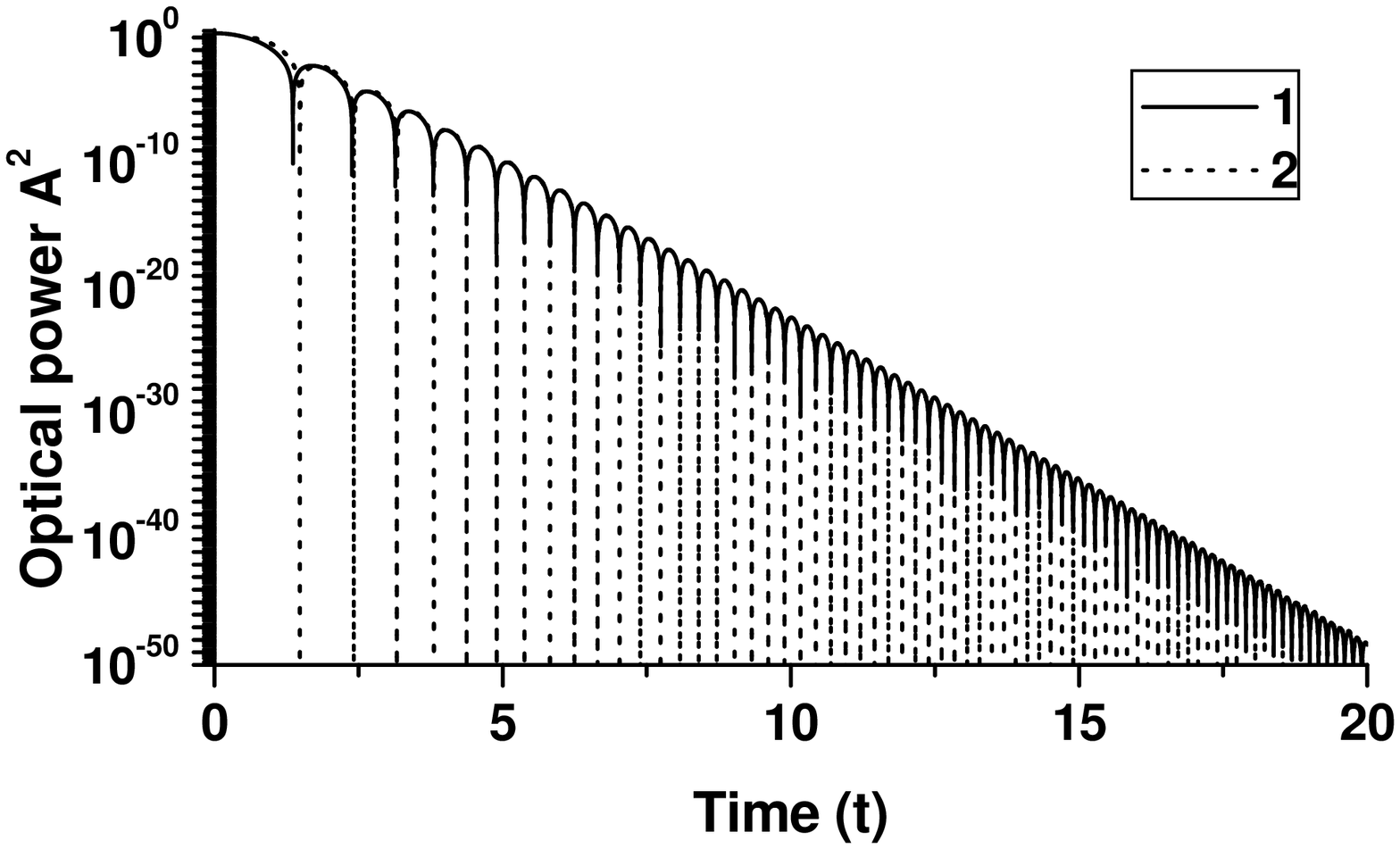}
\end{center}
{\bf (b)}
\begin{center}
\includegraphics[width = 3.5 in]{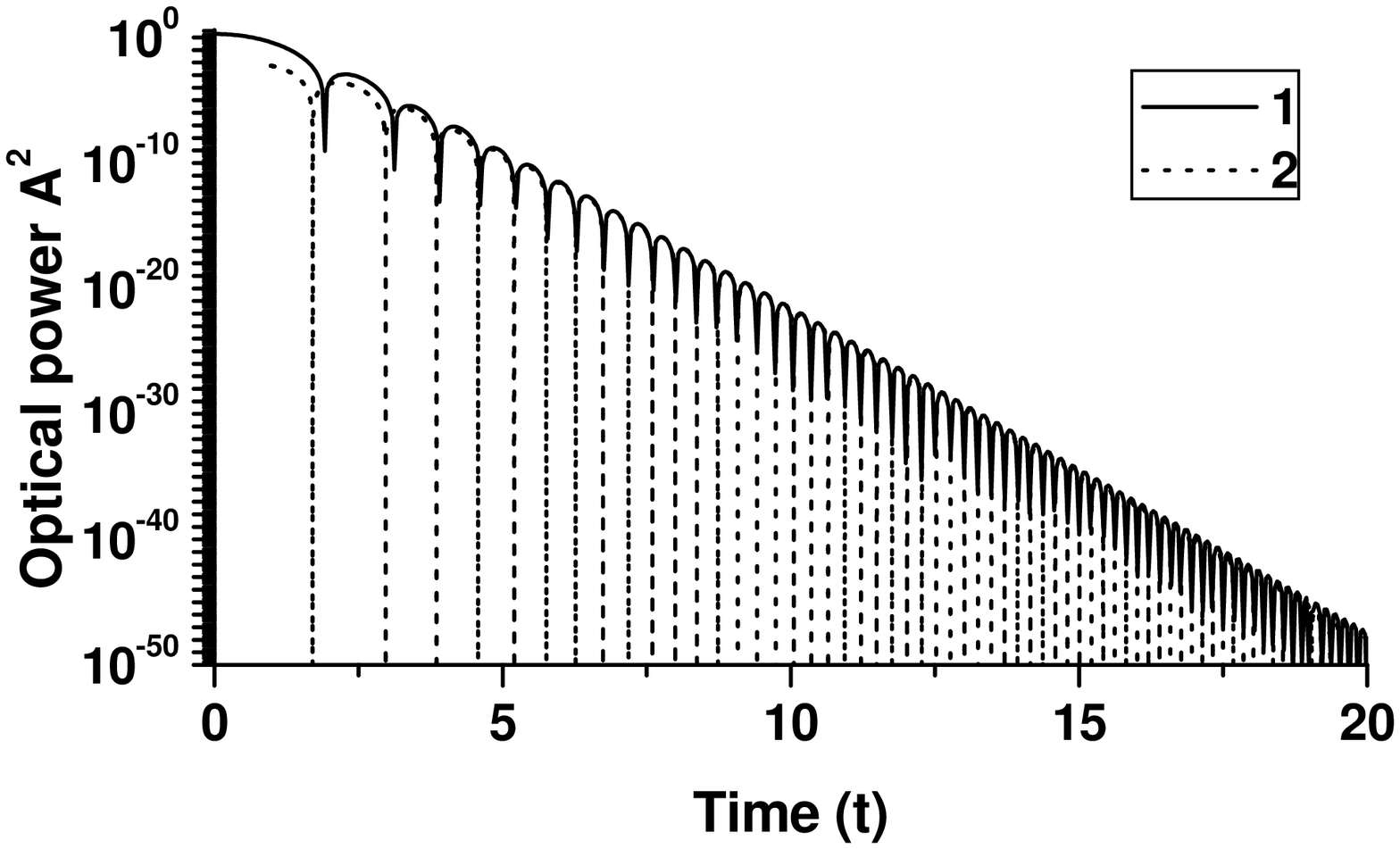}
\caption{DM soliton shape (curve 1) versus Eq. (4) (curve 2) for
(a) $\langle d\rangle=0,\ s=1, \ \lambda=1.$ and (b) for $\langle
d\rangle=0.1,\ s=1, \ \lambda=1.$ $A(t)$ is an
even function.} %\label{fig:fig2a}
\end{center}

\end{figure}
 high precision
numerical solution of Eq. $(\ref{psiint1c})$ for zero (Fig. 2a)
and nonzero (Fig. 2b) values of averaged dispersion.
 Numerical technique for solving Eq.
$(\ref{psiint1c})$ was developed in Refs.
\cite{lush2001a,lush2002a}. Analysis of numerical solutions in
Fig. 2 allows to make hypothesis that at leading order the
asymptotic behaviour of $A$ is given by
\begin{equation} \label{psishort0}
A_{asymp}(t) = f(t) \cos\{t^2[a_0+a(t)]\} \exp(-b |t|),
\end{equation}
where $a_0,\, b$ are constants and $f(t), \, a(t)$ are slow
functions of $t$. First it will be shown that an ansatz
$(\ref{psishort0})$ is consistent with Eq. $(\ref{psiint1c})$ at
leading order of $1/|t|$ provided $a_0=1/2s$ for $f(t)=const, \,
a(t)= const$ and later it will be found an expansion of $f(t), \,
a(t)$ over small parameter $1/|t|$. That result was announced
without derivation in Ref. \cite{lush2001a} for particular case
$\langle d\rangle=0$. Here general case which includes both
$\langle d\rangle=0$ and $\langle d\rangle\neq 0$ is considered.

First necessary condition for consistency of the ansatz
$(\ref{psishort0})$ with Eq. $(\ref{psiint1c})$ is that a
substitution of the envelope of soliton solution tails,
$A_{env}\sim \exp{(-b |t|)}$, into right hand side (r.h.s.) of Eq.
$(\ref{psiint1c})$ should recover the same exponential dependence
on $t$ after integration over $t_1, t_2$ variables in Eq.
$(\ref{psiint1c})$. Asymptotic behaviour for $|x|\to \infty$ of
kernel in r.h.s. of Eq. $(\ref{psiint1c})$ is given by:
\begin{equation} \label{ciasymp}
Ci(x)=sin(x)/x+O(1/x^2)
\end{equation}
and  $Ci(t_1t_2/s)$ produces at leading order the contribution to
phase of $A(t)$ only but not to envelope $A_{env}$. The
contribution of other factors in $r(t,t_1,t_2)$ to $A_{env}$ is
given by
\begin{eqnarray} \label{Acontr}
A(t_1+t)A(t_2+t)A(t_1+t_2+t)\nonumber \\
\simeq \exp\Big (-b
[|t_1+t|+|t_2+t|+|t_1+t_2+t|]\Big )
\end{eqnarray}
 and decays much faster than
$e^{-b|t|}$ outside the triangle ABC (shown in Fig. 3 for $t>0$).
\begin{figure}%[htbp]
\begin{center}
\includegraphics[width = 3.5 in]{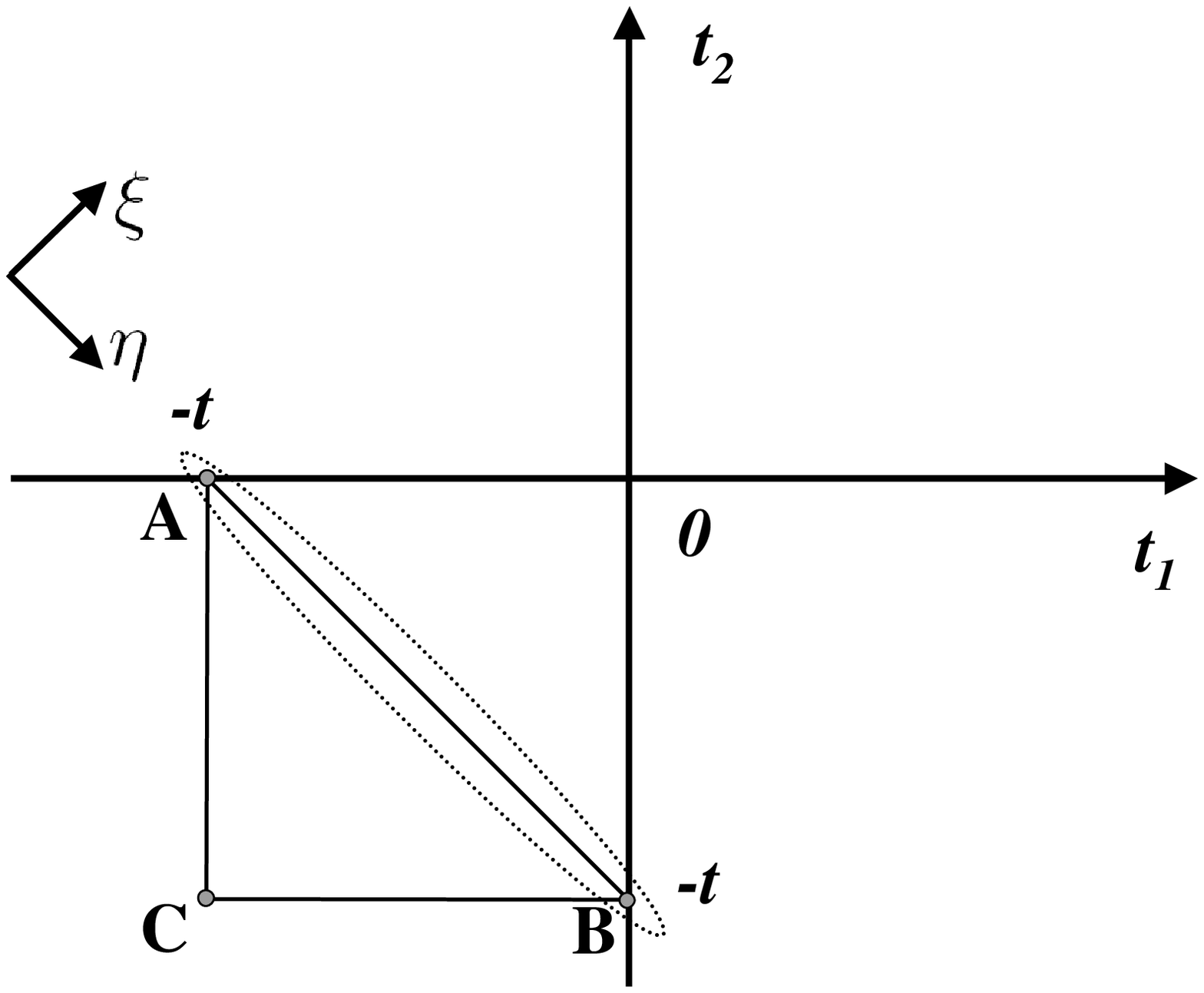}
%\vspace{0.5cm}
\caption{Integration plane $(t_1,t_2)$ for Eq. (3). $t>0.$
Characteristic layer is located along AB inside the dotted  curve.
lushnikovfig3.eps} %\label{fig:fig3}
\end{center}
\end{figure}
Triangle ABC is defined in $(t_1,t_2)$ plane and  determined from
condition $t_1+t>0, \ t_2+t>0, \ t_1+t_2+t<0$. However, inside
the triangle $A(t_1+t)A(t_2+t)A(t_1+t_2+t)\simeq e^{-b|t|}$,
which recovers the exponential decay as it appears at left hand
side (l.h.s.) of Eq. $(\ref{psiint1c})$. One can conclude that
$A_{env}\sim e^{-b |t|}$ is compatible with Eq.
$(\ref{psiint1c})$ and integration in plane $(t_1,t_2)$ inside an
area outlined by dashed curve in Fig. 3 gives leading order
contribution to r.h.s. of Eq. $(\ref{psiint1c})$ for $|t|\to
\infty$. The area inside the dashed curve includes triangle $ABC$
and characteristic layers of the width $O(t^0)$ which surround
that triangle. Integration over the area outside of dashed curve
in Fig. 3 gives exponentially small contribution of r.h.s. of Eq.
$(\ref{psiint1c})$. E.g. $A(t_1+t)A(t_2+t)A(t_1+t_2+t)\simeq
e^{-5b|t|}$ for $t_1\sim t/2, \ t_2\sim t/2$. This result is
immediately confirmed by numerical calculation of integral in
r.h.s of Eq. $(\ref{psiint1c})$ over the area inside the dashed
curve for DM soliton solution and comparing result of integration
with numerical integral over the whole $(t_1, t_2)$ plane.
Numerical integration was performed using numerical procedure of
Ref. \cite{lush2001a}. Note the important fact that exponential
envelope, $A_{env}$, is compatible with Eq. $(\ref{psiint1c})$
for any positive value of $b$ so that specific value of $b$,
corresponding to DM soliton can be only obtained from more detail
analysis of Eq. $(\ref{psiint1c})$.

Second necessary condition is that the substitution of fast
oscillating function $A_{osc}\sim \cos{(t^2[a_0+a(t)])}$ into
r.h.s. of Eq. $(\ref{psiint1c})$ should allow to reproduce the
same oscillations as in l.h.s. of Eq. $(\ref{psiint1c})$ for
$|t|\to \infty.$ Oscillating part of integrand $r(t,t_1,t_2)$ of
r.h.s. of Eq. $(\ref{psiint1c})$ consists of product of one $sin$
function (from asymptotic of oscillating kernel
$(\ref{ciasymp})$) and three $cos$ functions. Representing $sin$
and $cos$ through imaginary exponents one get a sum of 16 terms
with different values of argument. 14 of them are always
oscillate as a quadratic function of $t_1, \, t_2$ and thus give
vanishing contribution after integration over $t_1, \, t_2$. Rest
two terms can be made nonoscillating by appropriate choice of
$a_0$. Thus integrand $r(t,t_1,t_2)$ of r.h.s. of Eq.
$(\ref{psiint1c})$ at leading order, neglecting $a(t)$ inside the
triangle ABC, can be represented as $r(t,t_1,t_2)\sim \exp\Big (
it_1t_2/s+ia_0[(t_1+t)^2+(t_2+t)^2-(t_1+t_2+t)^2]-b|t| \Big
)-c.c. +f.o.=\exp( it_1t_2/s-2ia_0t_1t_2+ia_0t^2-b|t|)-c.c.
+f.o.$, where c.c. mean complex conjugation and  f.o. designates
14 fast oscillating terms. Oscillations (as a quadratic function
of $t_1, \, t_2$) of argument of two rest terms vanish identically
provided
\begin{equation}\label{a0}
 a_0=1/(2s),
\end{equation}
which is in excellent agreement with numerical solution of DM
soliton $(\ref{psiint1c})$. Next step is to take into account
nonzero value of $a(t)$ in Eq. $(\ref{psishort0})$. One can
series expand
\begin{equation}\label{at}
a(t)=a_1 /|t|+a_2/t^2+O(1/|t|^3)
\end{equation}
and find that term $a_1|t|$ in argument of $cos$ of Eq.
$(\ref{psishort0})$ causes oscillations of $r(t,t_1,t_2)$ as a
linear function of $t_1,t_2$. It is impossible to get rid of
these oscillations for any value of $a_1$ so one can conclude
that contribution of integration of $r(t,t_1,t_2)$ inside the
triangle ABC vanishes meaning that leading order contribution to
r.h.s of Eq.$(\ref{psiint1c})$ can only come from characteristic
layers of the width $O(t^0)$ which surround faces AB, BC and AC
of triangle ABC. In the neighborhood of each of these faces
either $t_1+t$ (face AC) or $t_2+t$ (face BC) or $t_1+t_2+t$
(face AB) is $\sim t^0$ so asympotic expression
$(\ref{psishort0})$ is not valid but rather exact solution of Eq.
$(\ref{psiint1c})$ should be used for respective factors
$A(t_1+t)$, $A(t_2+t)$ or $A(t_1+t_2+t)$ in $r(t,t_1,t_2)$. Still
there are fast oscillations of $r(t,t_1,t_2)$ as a function of
$t_1,t_2$ for integration inside the characteristic layers in
directions  parallel to faces BC and AC:
$A(t_1+t)A(t_1+t_2+t)\sim e^{ia_1
[2t+2t_1+t_2]-b|t_2|}+c.c.+f.o.$ (along BC) and
$A(t_2+t)A(t_1+t_2+t)\sim e^{ia_1
[2t+2t_1+t_2]-b|t_1|}+c.c.+f.o.$ (along AC). Thus integrations in
these 2 layers give a vanishing contribution and only an
integration in characteristic layer in the neighborhood of face
AB gives leading order contribution. To check that result the
numerical integration of integrand $r(t,t_1,t_2)$ with super
Gaussian modification factor: $\tilde r(t,t_1,t_2)\equiv
r(t,t_1,t_2)e^{-(|t+t_1+t_2|/t_0)^{(2n)}}$, $n\gg 1$, $1\ll
t_0\ll t$ was performed and it was found that $\int \tilde
r(t,t_1,t_2)dt_ dt_2=\int r(t,t_1,t_2)dt_ dt_2+O(1/|t|)$ which
confirms that only integration inside the characteristic layer
around AB (that layer is schematically shown in Fig. 3 as the area
inside the dashed curve) gives leading order contribution.

Analysis of an integration inside the characteristic layer AB is
convenient to be performed in new variables:
\begin{eqnarray}\label{xieta}
\xi = t+t_1+t_2, \\
\eta=t_1-t_2,
\end{eqnarray}
where axes $\xi$ and $\eta$ are perpendicular and parallel to
face AB, respectively (see Fig. 3). Face AB is located exactly at
axis $\xi=0$. Keeping in mind that the width of characteristic
layer is $\sim t^0\sim \xi$ while its length is $\sqrt{2} |t|\sim
\eta$ one gets that the ratio of typical values of $\xi$ and
$\eta$ is $\sim 1/|t|$ and $t_1+t=\frac{t+\eta}{2}+O(t^0), \
t_2+t=\frac{t-\eta}{2}+O(t^0), \ t_1=-\frac{t-\eta}{2}+O(t^0), \
t_2=-\frac{t+\eta}{2}+O(t^0).$ So that Eq. $(\ref{psiint1c})$,
using Eq. $(\ref{ciasymp})$, is reduced at leading order of
$1/|t|$ to
\begin{eqnarray}\label{gtred2}
\Big [\lambda g(\langle d \rangle)+\langle d \rangle t^2/s^2\Big ]
f(t) \cos{\big (\frac{t^2}{2 s}+a_1 |t|+a_2\big )}
\nonumber \\
\times \exp{(-b |t|)} =\int \frac{d\xi
}{4\pi}\int\limits^{t}_{0}d\eta \frac{A(\xi)
f(\frac{t-\eta}{2})f(\frac{t+\eta}{2})}{(\eta-t)(\eta+t)}
%\nonumber
\\ \times \sin{\Big ([\xi^2+t^2]/2 s+a_1 [\xi+|t|]+ 2a_2\Big )}
%\nonumber
%\\ \times
\exp{(-b \xi-b |t| )}, \nonumber
\end{eqnarray}
where the function $g(x)$ is used, $g(0)=1$ and $g(x) =1$ for
$x\neq 0$, because term with $\lambda$ in l.h.s. of Eq.
$(\ref{gtred2})$ is of leading order only if $\langle d
\rangle=0$. There is a separation of the integration variables
$\xi, \eta$ in r.h.s of Eq. and one can perform integration over
$\xi$  to get reduced equation:
\begin{eqnarray}\label{gtred3}
\Big [\lambda g(\langle d \rangle)+\langle d \rangle t^2/s^2\Big
] f(t) = C_1\int\limits^{t}_{0} d\tilde\eta \frac{
f(t-\tilde\eta)f(\tilde\eta)}{(t-\tilde \eta)\tilde\eta},
\end{eqnarray}
where $\tilde \eta =\frac{t+\eta}{2}$; it is assumed without loss
of generality that $t>0$; the real constant $C_1$ is given by
\begin{eqnarray} \label{C1}
C_1=\int \frac{d\xi}{8 \pi i}A(\xi)\exp{\Big[i(\xi^2/2 s+a_1 \xi+
a_2)-b \xi\Big ]}, \nonumber \\ C_1=Re(C_1),
\end{eqnarray}
and $C_1=Re(C_1)$ is the compatibility condition for Eq.
$(\ref{gtred2})$ meaning that phase of $cos$ in both r.h.s. and
l.h.s. (after integration over $\xi$) is the same and equals to
$a_2$ for $|t|=0$. It is important that exact (not asymptotic)
expression for $A(t)$ should be used in evaluating integral in
Eq. $(\ref{C1})$ which means that asymptotic tails
$(\ref{psishort0})$ are determined by strongly nonlocal effects
because all time scales contribute to $C_1$.

The integral term in Eq. $(\ref{gtred3})$ is the convolution
integral and using Laplace transform, $\tilde f_p\equiv
\int^\infty_0 dt \tilde f(t)e^{-pt}$, for function $\tilde
f(t)=f(t)/t$ one gets an ordinary differential equation:
\begin{equation}
 \label{lapl1}
\lambda g(\langle d \rangle)\frac{d \tilde f_p}{d p}+\langle d
\rangle \frac{1}{s^2}\frac{d^3 \tilde f_p}{d p^3}= -C_1\tilde
f_p^2,
\end{equation}
which has the following solution: $\tilde f_p=g(\langle d \rangle)
\frac{\lambda}{C_1(p+C_2)}+60\frac{\langle d
\rangle}{C_1(p+C_2)^3}$, where $C_2$ is an arbitrary real
constant. Inversion of Laplace transform results in: $f(t)=\Big [
tg(\langle d \rangle) \lambda+30 t^3\langle d \rangle\Big ]e^{-C_2
t}/C_1$. For arbitrary sign of $t$ one should replace $t$ by
$|t|$ in this expression. Factor $e^{-C_2 |t|}$ gives
renormalization of constant $b$ in Eq. $(\ref{psishort0})$, which
reflects a freedom of choice of $b$ in Eq. $(\ref{psishort0})$ so
$C_2$ could be set, without loss of generality, to zero and
expression for $f(t)$ takes the following final form:
\begin{equation}
 \label{fres1}
f(t)=\frac{\lambda |t|}{C_1} \ \ \mbox{for} \ \ \langle d
\rangle=0, \quad f(t)=\frac{30\lambda |t|^3}{C_1} \ \ \mbox{for} \
\ \langle d\rangle \neq 0.
\end{equation}

Eqs. $(\ref{at}),(\ref{fres1})$ have four real constants $C_1, \,
b,\, a_1$ and $a_2$ which can not be determined from leading
order expansion developed in this Article because all time scales
contribute to numerical values of these constants. One can easily
find constants $C_1, \, b,\, a_1, \, a_2$ from a fit with
numerical solution of Eq. $(\ref{psiint1c})$. Phase of $cos$ in
Eq. $(\ref{psishort0})$ is determined from zeros of the numerical
solution of $A(t)$ keeping in mind that coefficient $a_0=1/2s$ is
known analytically.  After the constants $a_1, \, a_2$ are found,
one can  determine $C_1$ and $b$ numerically from a two-parameter
fit with envelope of DM soliton tail oscillations.
 Fig. 2a,b show a good agreement between the numerical solution
 (curve 1) and $A_{asymp}(t)^2$
dependence (curve 2). Fig. 2a corresponds to $s=1, \lambda=1,
\langle d\rangle=0$  and numerical values $C_1=0.08357,\,
b=3.0452,\, a_1=1.4136,\, a_2=1.5102$. Fig. 2b corresponds to
$s=1, \lambda=1, \langle d\rangle=0.1$ and numerical values
$C_1=13.818,\, b=3.2483,\, a_1=0.1443,\, a_2=-0.1276$. $C_1, \,
b, \, a_1,\,a_2$ are obtained from the above described fitting
procedure. A similar plot for zero average dispersion $\langle
d\rangle=0$ was given in Ref. \cite{lush2001a} in Fig. 3. Note
however that curve 2 in Fig. 2b converges to curve 1 for larger
$t$ compare to the case $\langle d\rangle=0$ in Fig. 2a which
reflects the fact that for intermediate values of time, $t\sim
t_i$, $\langle d\rangle t_i^2= \lambda$, there is a transition
from the solution of $f(t)$ in Eq. $(\ref{fres1})$ for $\langle
d\rangle=0$ to the solution for $\langle d\rangle\neq 0.$ If
$\langle d\rangle$ is large enough
 so that $t_i$ $\sim$ DM soliton width, then there is a
transition from oscillating tails of DM soliton to nonoscillating
exponential tails of NLS with constant dispersion $\langle
d\rangle$. NLS soliton width, $(\lambda/\langle d\rangle)^{1/2},$
 is much larger than DM soliton width for small $\langle d\rangle$.

\section{Conclusion}

In summary, the asymptotic, $|t|\to \infty$, of DM soliton tails
is given by Eqs.
$(\ref{psishort0}),(\ref{a0}),(\ref{at}),(\ref{fres1})$ which is
in agreement with numerical solution of Eq. $(\ref{psiint1c})$.
 The pre-exponential factor (\ref{fres1}) is qualitatively different
for zero and nonzero path-averaged dispersion. An exponential
decay of DM soliton envelope, which is much slower compare to
Gaussian-type decay near soliton center, suggests that interaction
of different DM solitons is strong and could affect bit-rate
capacity of optical lines. A detailed investigation interaction
between DM solitons is outside the scope of this Article. Another
possible direction of future research is to study the effect of
losses on DM soliton tails as well as DM soliton with nonlinearity
compensation \cite{GabitovLushnikov2002}. Qualitatively
oscillations are still of the form $(\ref{psishort0})$ but
further research is needed.

The author thanks M. Chertkov, I.R. Gabitov, E.A. Kuznetsov, V.E.
Zakharov and V. Zharnitsky for helpful discussions. The support
was provided by  the Department of Energy, under contract
W-7405-ENG-36. The author's E-mail address is
lushnikov@cnls.lanl.gov.

%\newpage

~
\newpage

\end{document}